\documentclass[a4paper]{article}
\usepackage{multirow}
\usepackage{subfigure}
\usepackage{INTERSPEECH2022}
\usepackage{hyperref}
\hypersetup{
    colorlinks=true,
    linkcolor=blue,
    filecolor=magenta,      
    urlcolor=cyan,
    pdftitle={Overleaf Example},
    pdfpagemode=FullScreen,
    }

\title{Cross-Age Speaker Verification: Learning Age-Invariant Speaker Embeddings }
\name{Xiaoyi Qin$^{1,2}$, Na Li$^{3}$, Chao Weng$^{3}$, Dan Su$^{3}$, Ming Li$^{1,2}$\thanks{Correspond author: Ming Li. This research is funded in part by the National Natural Science Foundation of China (62171207), Tencent AI Lab Rhino-Bird Gift Fund and Science and Technology Program of Guangzhou City (202007030011) . Many thanks for the
computational resource provided by the Advanced Computing East China Sub-Center.}}

\address{$^{1}$School of Computer Science, Wuhan University, Wuhan, China \\
		$^{2}$Data Science Research Center, Duke Kunshan University, Kunshan, China\\
		$^{3}$Tencent AI Lab, Shenzhen, China}
\email{ming.li@whu.edu.cn}

\begin{document}

\maketitle

\begin{abstract}
Automatic speaker verification has achieved remarkable progress in recent years. However, there is little research on cross-age speaker verification (CASV) due to insufficient relevant data. In this paper, we mine cross-age test sets based on the VoxCeleb dataset and propose our age-invariant speaker representation(AISR) learning method. Since the VoxCeleb is collected from the YouTube platform, the dataset consists of cross-age data inherently. However, the meta-data does not contain the speaker age label. Therefore, we adopt the face age estimation method to predict the speaker age value from the associated visual data, then label the audio recording with the estimated age. We construct multiple Cross-Age test sets on VoxCeleb (Vox-CA), which deliberately select the positive trials with large age-gap. Also, the effect of nationality and gender is considered in selecting negative pairs to align with Vox-H cases. The baseline system performance drops from 1.939\% EER on the Vox-H test set to 10.419\% on the Vox-CA20 test set, which indicates how difficult the cross-age scenario is. Consequently, we propose an age-decoupling adversarial learning (ADAL) method to alleviate the negative effect of the age gap and reduce intra-class variance. Our method outperforms the baseline system by over 10\% related EER reduction on the Vox-CA20 test set. The source code and trial resources are available on \href{https://github.com/qinxiaoyi/Cross-Age\_Speaker\_Verification}{https://github.com/qinxiaoyi/Cross-Age\_Speaker\_Verification}.

\end{abstract}
\noindent\textbf{Index Terms}: speaker verification, cross-age, age-invariant

\section{Introduction}
\label{sec.intro}

Speaker verification has achieved great success thanks to the deep learning application. Automatic speaker verification (ASV) using X-vector\cite{xvector} with its variant\cite{ecapatdnn,ftdnn,cai_exploring_2018} can extract a fix-dimensional discriminative identity vector from the variable-length audio recording. The margin-based loss functions\cite{sphereface,arcface} are also adopted to train the speaker verification system with large-scale databases to reduce the intra-speaker variability and increase the inter-speaker distance. Despite the remarkable success of ASV, how to handle the cross-age variability is still a challenge. In addition, there are many practical applications for cross-age speaker verification (CASV), such as identifying the suspect from telephone fraud audio recorded long term ago. To the best of our knowledge, the related study is rare due to insufficient relevant data. Early researches \cite{modadp_sv,sv_longtime_aging,effect_ca_sv,voiceaging_sr} focus on the small-scale datasets since the collection of cross-age speech data is a long and expensive process. Lei et.al\cite{roleage_sv} analyze the influence of age on the NIST SRE08 data. Sean et.al \cite{AISE_diarizaion} propose an age-invariant speaker embedding approach to speaker diarization. However, the evaluation sets on those datasets do not contain the cross-age speaker verification scenario. Some researches \cite{age_nist_ivector,age_nist_e2e,age_timit,age_ml,age_ivector_ml} also experiment on the NIST SRE and TIMIT\cite{timit} datasets to estimate speaker age. Unfortunately, each speaker's data only cover one age phase. Fenu et.al \cite{fairvoice} studies the impact of demographic imbalance on group fairness in speaker recognition and also consider the age influence. But, no large-scale cross-age dataset is available for cross-age speaker verification. 

VoxCeleb dataset \cite{vox1,vox2} has been widely used as the benchmark for general speaker verification. Since the VoxCeleb dataset is collected from the YouTube platform, the dataset consists of cross-age data inherently. However, there is no age label in the meta-data. \cite{agevoxceleb} and \cite{vox_enrichment}  provide partial VoxCeleb2 age labels for the speaker age estimation task; age label is estimated using celebrities' birth year and manual annotation on the year when the video is recorded. However, manual labeling is time-consuming and may not apply to most data for large-sacle ASV tasks. Therefore, we adopt the face age estimation method to automatically obtain the face age value and conduct age labeling for all audio of VoxCeleb1 and VoxCeleb2. Although the labeled age value is not completely accurate, it can rank all the audio of one speaker in chronological order. Thus, we propose a cross-age speaker verification task and construct multiple Cross-Age test sets on VoxCeleb (Vox-CA), which deliberately selects the positive trials with large age gap. In addition, as mentioned in the VoxCeleb-H test set, the negative pair also take the effect of nationality and gender into account. The system results on Vox-CA confirm how difficult the cross-age scenario is. 

The cross-age scenario is challenging because the aging process indeed enlarges the intra-class variance. The age-invariant representation learning is a hot topic in face recognition \cite{calfw,dal_aifr,orthogonal_decom_aifr,aifr_fas}. Inspired by those works, we propose the Age Decoupling Adversarial Learning (ADAL) module to encourage speaker identity features to have smaller intra-class variations and tend to be less correlated with the age information. The proposed ADAL approach uses an attention mechanism to extract age-related information from high-level feature maps and decouples the age component from the speaker embedding. A margin-based identity classifier is adopted to model the residual identity feature. The adversarial learning based age classification also operates on residual identity features to weaken the remaining age information. The proposed method helps the samples from the same speaker with different ages be aggregated into one center to derive the age-invariant speaker embedding. The final result outperforms the state-of-the-art baseline system.

The rest of this paper is organized as follows. In section \ref{sec.voxca}, we describe the details of Vox-CA and compare it with the VoxCeleb1 test set. The proposed ADAL method is introduce in Section \ref{sec.dal}. Section \ref{sec.experiment} reports the experimental results on Vox-CA. Conclusions are provided in Section \ref{sec.conclu}.

\section{VoxCeleb Cross-Age Test Set}
\label{sec.voxca}  
\vspace{-0.1cm}
\begin{figure}[htbp]
	\centering
	
	\begin{minipage}[t]{0.999\linewidth}
	\centering
		\includegraphics[width=1.6in]{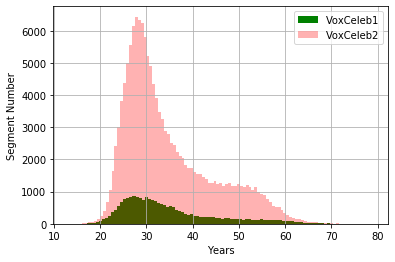}
		\caption{Segment age distribution of VoxCeleb}
		\label{fig:age_dis}
	\end{minipage}
	\begin{minipage}[t]{0.999\linewidth}
	\centering
		\includegraphics[width=1.6in]{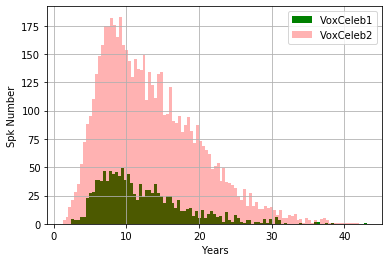}
		\caption{Distribution of maximum age gap to VoxCeleb}
		\label{fig:age_gap}
	\end{minipage}
\end{figure}
\vspace{-0.5cm}
\subsection{Related Work}

VoxCeleb1 test set is a benchmark used for speaker verification, which consists of the original VoxCeleb1 test set (Vox-O), the VoxCeleb1-E test set (Vox-E), and the VoxCeleb1-H test set (Vox-H). Vox-E is an evaluation protocol covering the entire dataset with 1,251 speakers. Vox-H is another evaluation protocol in which all negative pairs are from the same nationality and gender. We construct the Cross-Age test sets on VoxCeleb, named Vox-CA.

\subsection{Construction Details}
The construction pipelines adopt the following steps:
\begin{itemize}
\item Gathering the face image from meta-data of VoxCeleb1\footnote{\href{https://www.robots.ox.ac.uk/~vgg/research/CMBiometrics/}{https://www.robots.ox.ac.uk/$\sim$vgg/research/CMBiometrics/}} and VoxCeleb2\footnote{\href{https://www.robots.ox.ac.uk/~vgg/data/voxceleb/vox2.html}{https://www.robots.ox.ac.uk/$\sim$vgg/data/voxceleb/vox2.html}}.

\item Estimating the age of each face image.
\item Labeling the estimated age value for each audio utterance.
\item Selecting large age-gap audios as positive pairs and the pairs of same nationality and gender  as negative pair 
\end{itemize}

For the sake of clarity, the key stages are described in the following paragraphs:

\noindent \textbf{Estimating and labeling age for audio}. We use the Dex \cite{dex}, the winner of the ChaLearn LAP2015 challenge, to estimate the age for each face image. Since each utterance audio corresponds to multiple face images, the average age value of faces is used as the estimated age for this utterance. In addition, as the utterance audio is split from the video segment, the speaker file structure is formed by: 
\texttt{speaker\_id/segment\_id/utterance\_id.wav}.
Thus, the related utterances of the same segment should share the same age. The segment age, the average age among all the utterances belonging to the same segment, is determined as the final age label. The age distribution is shown in Figure.\ref{fig:age_dis}. Most of the age values are in the range between 20 and 70. 

\noindent \textbf{Forming positive/negative pairs}.
Since the VoxCeleb2 dataset is adopted for training under most ASV implementation scenarios, the entire VoxCeleb1 dataset contributes to the construction of the cross-age test set with the following rules.

First, the positive pairs must be the cross-age case. In other words, the pair audios can not be from the same video segment. We count the maximum age gap of each speaker and present the distribution in Figure.\ref{fig:age_gap}. It is observed that the largest age gap for most speakers is between 0 and 20 years. The test set should cover at least 80 speakers to be meaningful. However, especially for VoxCeleb1, far fewer speakers have more than 20 years of age-gap data. Thus, the number of enrollment speakers should also be considered when constructing the test set. 

Second, following the Vox-H setting, all negative pairs are constructed within the same nationality and gender. We keep the same setting with Vox-H that each nationality-gender combination has 5 individuals at least.   
    
Following the rules mentioned above, there are four Vox-CA sets are constructed according to different age-gap categories:
\begin{itemize}
	\item Vox-CA5. The age gap of the positive pair is  5 years at least. The candidate speakers must possess more than 7 years of max age-gap data.
	\item Vox-CA10. The age gap of the positive pair is  10 years at least. The candidate speakers must possess more than 12 years of max age-gap data.
	\item Vox-CA15. The age gap of the positive pair is  15 years at least. The candidate speakers must possess more than 17 years of max age-gap data. 
	\item Vox-CA20. The age gap of the positive pair is  20 years at least. The candidate speakers must possess more than 22 years of max age-gap data.
\end{itemize}

The Vox-CA proposes a challenging task covering the cross-age, same nationality and same gender cases. In addition, we also implement the single variable test set, including but not limited:1) test set within the cross-age; 2) test set within the same nation; 3) test set within the same gender; 4) test set within the intra-segment, to observe the effect of various factors on verification. The results are reported in Section.\ref{sec.experiment}.

\begin{table}[htbp]\centering \footnotesize
    \caption{\label{tab:intra-segment} {The performances of baseline system on the VoxCeleb1 test set with or without the intra-segment positive pairs. 
}}
    \begin{tabular}{lccc}
    \toprule
    \textbf{Test set} & \textbf{Intra-segment} & \textbf{EER[\%]} & \textbf{minDCT}  \\
    \midrule
    Vox-O & yes & 0.962 & 0.100   \\ 
    Vox-O & no  & 1.027 & 0.109   \\
    Vox-E & yes & 1.094 & 0.122   \\ 
    Vox-E & no  & 1.173 & 0.132   \\
    Vox-H & yes & 1.939 & 0.200   \\ 
    Vox-H & no & 2.078 & 0.217   \\ 
    \bottomrule
    \end{tabular}
    \end{table}
    \vspace{-0.5cm}

\subsection{Comparison of Vox-E, Vox-H and Vox-CA}

In this part, we compare the difference of Vox-E, Vox-H and Vox-CA from the following aspects.

\textbf{Positive pair within intra-segment}. In this case, the pair audios are selected from the same video segment. Therefore, the successful verification rate is higher. We count the intra-segment pairs from the VoxCeleb1 test set and re-calculate the performance after removing those pairs. The results are reported in Table.\ref{tab:intra-segment}. The results indicate that intra-segment cases could improve the performance but do not reflect the real application. In Vox-CA, the positive audio pairs are compulsorily selected from different segments.

\textbf{Positive pair within the cross-age}. From Table.\ref{tab:trial}, the average age-gap of VoxCeleb1 test set is about 3 years. The positive pairs of the VoxCeleb1 test set are randomly chosen from the same person without considering the age gap. The Vox-CA selects pair audio from large age-gap segments.

\textbf{Negative pair within the same nationality and gender}. Both Vox-H and Vox-CA take nationality and gender into account when constructing negative pairs. The pairs of Vox-E are randomly selected from the entire dataset. Thus, the Vox-H and Vox-CA sets are more challenging.   

\begin{table}[htbp]\centering \scriptsize 
    \caption{The statistics of the VoxCeleb1 test set and Vox-CA. \textit{Trials Num.} and \textit{Spk. Num.} describe the number of trials and enrollment speakers, respectively. The column of \textit{positive} and \textit{negative} present the mean and standard deviation of the age-gap values in the corresponding pairs.}
     \label{tab:trial}
    \begin{tabular}{lccccc}
    \toprule
	\multirow{2}*{\textbf{Test set}} & \multirow{2}*{\textbf{Spk. Num.}} & \multirow{2}*{\textbf{Trials Num.}} & \multicolumn{2}{c}{\textbf{Age-gap}} \\
	\cmidrule(lr){4-5} & & & \textbf{Positive} & \textbf{Negative} \\
    \midrule 
    Vox-O  & 40   & 37611  & 2.68 $\pm$ 2.88 & 15.50 $\pm$ 12.46 \\
    Vox-E  & 1251 & 579818  & 3.14 $\pm$ 3.48 & 12.05 $\pm$ 9.81 \\
    Vox-H  & 1190 & 550894  & 3.14 $\pm$ 3.47 & 11.27 $\pm$ 9.42 \\
     
    \midrule
	Vox-CA5  & 971 & 370540 &  9.98 $\pm$ 3.94 & 12.36 $\pm$ 9.58 \\
	Vox-CA10 & 506 & 151384 &  15.29 $\pm$ 3.44 & 14.66 $\pm$ 9.93 \\
	Vox-CA15 & 215 & 54608  &  20.39 $\pm$ 3.38 & 16.63 $\pm$ 10.24 \\
	Vox-CA20 & 85  & 18888  &  25.28 $\pm$ 2.87 & 18.42 $\pm$ 10.58 \\
    \bottomrule
    
    \end{tabular}
    \end{table}
    \vspace{-0.5cm}

\section{Learning Age-invariant Speaker Embeddings}  
\label{sec.dal}

\subsection{Decouple Age-related Component}
\begin{figure}
  \centering
  \includegraphics[width=\linewidth]{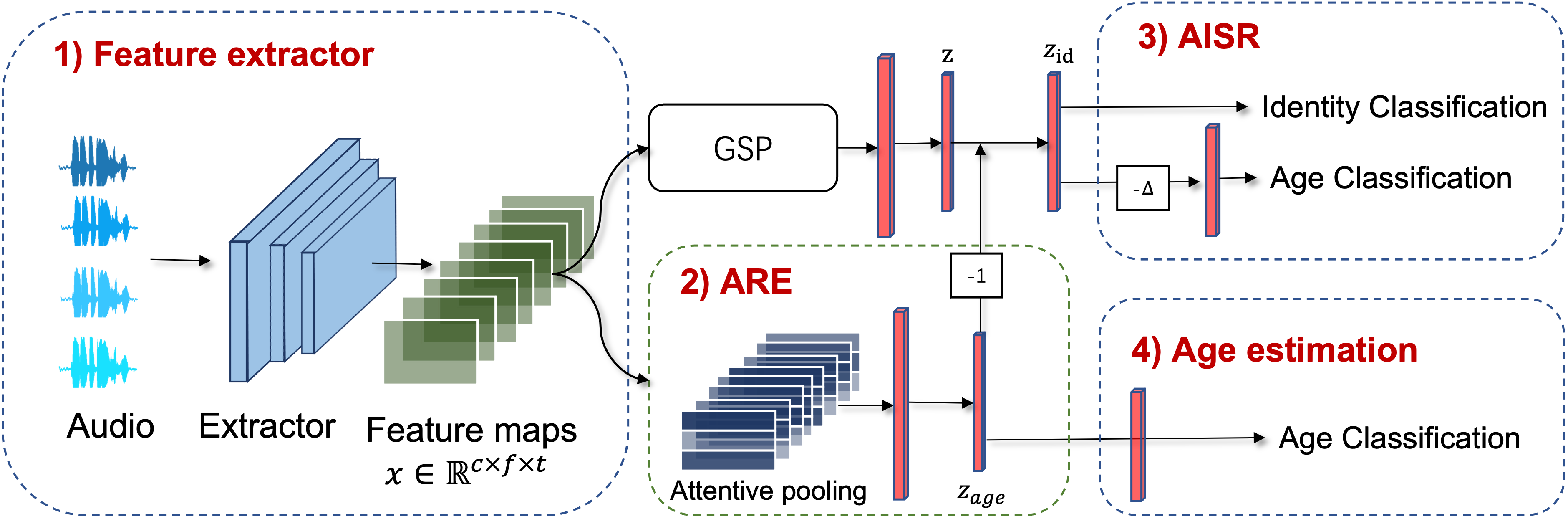}
  \caption{An overview of the proposed ADAL structure. The AISR denotes the Age-Invariant Speaker Representation. }
  \label{fig:network}
  \vspace{-0.4cm}
\end{figure}

Feature embedding consists of identity information and age information as driven by the two corresponding tasks. Motivated by this, we design a linear model to decouple age information from identity features. First, we assume that feature embedding $\bf{z}\in \mathcal{R}^d$, a $d$ dimensional vector extracted from an input audio $I$, is the sum of identity and age information:
\begin{equation}
	\bf{z} = \bf{z}_{id}+\bf{z}_{age}
\end{equation}
where the $\bf{z}_{id}$ and $\bf{z}_{age}$ denote that identity component and age component, respectively. Then we design an age-related extractor module (ARE), which uses the attention mechanism to extract age-related information from the high-level feature maps $\bf{x}\in \mathcal{R}^{c\times f \times t}$. Then, following the pooling and linear layer, we get the age-related embedding with $d$ dimensions, which can be formulated as follows::
\begin{equation}
	\bf{z}_{age} = ARE(\bf{x}) \\
            = fc(pool(x*\sigma(\bf{x})))
\end{equation}
where $\sigma$ denotes the attention module, $pool$ and $fc$ represent the pooling layer and fully connection layer. By subtracting $\bf{z}_{age}$ from $\bf{z}$,, the age-related component supervised by an age classifier is peeled off from feature embedding. Figure.\ref{fig:network} present the details of the whole network structure. In this paper, we adopt the Attentive Statistical Pooling (ASP) \cite{asp_pooling} to highlight the age-related information from high-level feature map and obtain the fix-length age embedding vector.   

\vspace{-0.32cm}
  
\subsection{Multi-task Learning}
As shown in Figure.\ref{fig:network}, the model consists of three supervised tasks: identity classification, age classification, and age adversarial learning.

\textbf{Identity classification}. We adopt the identity classifier $C$ to guide the $z_{id}$ to represent the identity information. As speaker timbre can change considerably over time, the essential problem of CASV is that speaker aging leads to increasing intra-class variance. Therefore, we adopt ArcFace as an identity loss function to reduce the intra-class distance.

\textbf{Age classification}. To encourage the age information to be decoupled from the speaker embedding, an age group classifier $A$ is leveraged to supervise the age-related embedding learning. However, we do not use age regression or age classifier as ground truth in common speaker age estimation methods used. Since the age value is estimated by Dex\cite{dex} but not the ground truth, the age labels are noisy. Therefore, we adopt the age group classifier instead, age is split into 7 groups: 0-20,21-30,31-40,41-50,51-60,61-70 and 70-100.    

\textbf{Age adversarial learning}
To make sure the identity embedding $\bf{z}_{id}$ is towards age-invariant, an additional age classifier with gradient reversal layer (GRL) \cite{grl} is applied upon the $\bf{z}_{id}$ to reduce the residual age information further. 

Thus, the method is named as Age Decoupling Adversarial Learning (ADAL). The final loss for the method is formulated as:
\vspace{-0.2cm}
\begin{gather}
	\mathcal{L}_{id}(\bf{z}_{id})=l_{ce}(C(\bf{z}_{id}),y_{id}) \\
	\mathcal{L}_{age}(\bf{z}_{age})=l_{ce}(A(\bf{z}_{age}),y_{age}) \\
	\mathcal{L} = \mathcal{L}_{id}(\bf{z}_{id})+\lambda_{age} \mathcal{L}_{age}(\bf{z}_{age})+\lambda_{grl}\mathcal{L}_{age}(GRL(\bf{z}_{id}))
\end{gather}
\vspace{-0.2cm}

where $y_{id}\in \{0,1,\dots,N\}$ and $y_{age}\in \{0,1,\dots,6\}$ are the output labels of identity and age estimation task, respectively. $l_{ce}$ denotes the cross-entropy loss, $\lambda_{age}$ and $\lambda_{grl}$ are scalars to balance different loss terms. 

\begin{table*}[tp] \scriptsize 
  \caption{The performance of different speaker verification systems in terms of EER. The model with GRL describes the simplest adversarial learning that uses GRL upon the $\bf{z}$ vector to perform the age classification task, which makes the speaker embedding less correlated to age. In the Age Residual method, $\bf{z}_{id}$ is the residual part between $\bf{z}$ and $\bf{z}_{age}$, the $\bf{z}_{age}$ is extracted from $\bf{z}$ and supervised by age classification. The model equipped with ARE represents the $\bf{z}_{age}$ is also supervised by age classification but extracted by ARE.}

  \label{tab:daal_result}
  \centering
  \begin{tabular}[c]{lccccccccccccccccc}
    \toprule
 	\multirow{2}*{\textbf{Model}} & \multirow{2}*{\textbf{Vox-E}} & \multirow{2}*{\textbf{Vox-H}} & \multicolumn{4}{c}{\textbf{Cross-age}} &  \multicolumn{4}{c}{\textbf{Cross-age \& Same nationality \& Same gender}}\\
  \cmidrule(lr){4-7} \cmidrule(lr){8-11} & & & \textbf{Only-CA5} & \textbf{Only-CA10} & \textbf{Only-CA15} &  \textbf{Only-CA20} &\textbf{Vox-CA5}   & \textbf{Vox-CA10} &\textbf{Vox-CA15}   & \textbf{Vox-CA20}\\

   \midrule
   	 ResNet34-Softmax  & 2.798\% & 4.806\% & 4.310\% & 6.004\% & 8.019\% & 9.308\% & 7.366\% & 9.215\% & 12.405\% & 14.888\% \\
	 ResNet34-ArcFace  & \bf{1.094\%} & 1.939\% & \bf{1.953\%} & 3.437\% & 5.927\% & 8.185\% & 3.407\% & 4.974\% & 8.028\% & 10.419\% \\
	 \quad + GRL & 1.122\% & \bf{1.934\%} & 2.021\% & 3.579\% & 6.036\% & 8.566\% & \bf{3.405\%} & 4.949\% & 8.017\% & 10.610\% \\
	 \quad + Age Residual & 1.121\% & 1.960\% & 2.040\% & 3.536\% & 5.871\% & 7.864\% & 3.499\% & 5.078\% & 8.039\% & 10.229\% \\
	 \quad + ARE (\bf{ours}) & 1.108\% & 1.951\% & 1.980\% & 3.345\% & 5.719\% & 7.803\% & 3.431\% & \bf{4.814\%} & 7.786\% & 9.911\% \\
	 \quad + ADAL (\bf{ours})& 1.121\% & 1.974\% & 1.991\% & \bf{3.330\%} & \bf{5.540\%} & \bf{7.442\%} & 3.441\% & 4.822\% & \bf{7.515\%} & \bf{9.519\%} \\
     \bottomrule
     \end{tabular}
     \vspace{-0.32cm}
\end{table*}
  
 \begin{table}[!htbp]\centering \footnotesize
    \caption{\label{tab:trial_result} {Results on different test sets based on the ResNet-GSP-ArcFace model.}}
    \begin{tabular}{lccc}
    \toprule
    \textbf{Test set} & \textbf{Construct} & \textbf{EER[\%]} & \textbf{mDCF$_{0.01}$}  \\
    \midrule
   	\bf{Vox official} \\
    Vox-O & random  & 0.962\% & 0.100   \\ 
    Vox-E & random  & 1.094\% & 0.122   \\ 
    Vox-H & nation\&gender  & 1.939\% & 0.200   \\ 
    \midrule
   	\bf{our proposed} \\
   	our-E     & random  & 1.202\% & 0.123 \\ 
    our-H     & nation \& gender  & 2.044\% & 0.192 \\
    only-N    & nation  & 1.568\% &  0.164 \\ 
    only-G    & gender & 1.534\%  & 0.146   \\
    only-I    & intra-segment & 0.227\%  & 0.015   \\
    only-CA5  & age     &  1.953\% & 0.177     \\
    only-CA10 & age     & 3.437\% & 0.272    \\
    only-CA15 & age     & 5.927\% & 0.352   \\
    only-CA20 & age     & 8.185\% & 0.464   \\ 
    Vox-CA5 & age \& nation \& gender  & 3.407\% & 0.300   \\ 
    Vox-CA10 & age \& nation \& gender  & 4.974\% & 0.370   \\ 
	Vox-CA15 & age \& nation \& gender  & 8.028\% & 0.481   \\ 
	Vox-CA20 & age \& nation \& gender  & 10.419\% & 0.646   \\ 

    \bottomrule
    \end{tabular}
    \vspace{-0.32cm}
    \end{table} 
    \vspace{-0.2cm}
  
\section{Experimental Results}
\label{sec.experiment}
\subsection{Implementation Details}
\textbf{Network}. For the baseline system, we adopt the ResNet34\cite{resnet} as the backbone. The widths (channels number) of the residual blocks are \{32, 64, 128, 256\}. The global statistic pooling (GSP) layer, which computes the mean and standard deviation of the output feature maps, can project the variable length input to the fixed-length vector. The output of a fully connected layer with 128 dim followed after the pooling layer is adopted as the speaker embedding layer. The ArcFace-based classifier \cite{arcface} (s=64,m=0.2), which increase intra-speaker distances while ensuring inter-speaker compactness, is used to the identity task. In addition, we also provide the Softmax classifier as a comparison. For the ADAL method, the ASP is adopted as ARE module to extract the $\bf{z}_{age}$ vector. For the age classification, we stack the FC-ReLU-FC structure upon the $\bf{z}_{age}$ and $\bf{z}_{id}$ to predict the age group value. 

\textbf{Data Processing}. The acoustic features are 80-dimensional log Mel-filterbank energies with a frame length of 25ms and hop size of 10ms. We adopt the on-the-fly data augmentation \cite{on-the-fly} to diversify training samples. Four types of augmentation methods we adopted: 1) adding noise using MUSAN \cite{musan} dataset; 2) adding convolutional reverberation using RIR Noise \cite{RIR} datasets; 3) changing amplification, and 4) changing audio speed (pitch remains untouched).

\textbf{Training Details}. The SGD optimizer is employed to update the model parameters. We adopt the multi-step learning rate(LR) scheduler with $0.1$ initial LR; the decay step and factor are 10 and 0.1, respectively. We adopt the linear warmup from $0.0$ to $0.1$ LR in the first two epochs to prevent the training instability and speed model convergence. Training stopped after LR dropped to 1e-5. The hyper-parameters in loss are set as following: $\lambda_{age}=0.1$ and $\lambda_{grl}=0.1$.

\textbf{Evaluation Measures}. Cosine similarity is used for trial scoring. Verification performances are measured by EER and the minimum normalized detection cost function (mDCF) with $P_\mathrm{target} = 10^{-2}$ and $C_\mathrm{FA} = C_\mathrm{Miss} = 1$.

\subsection{Experimental results of the baseline method} 
In this part, we introduce the ResNet34-GSP-ArcFace system for speaker verification. We compare the baseline performance on Vox-O, Vox-E, Vox-H and our proposed Vox-CA. Table.\ref{tab:trial_result} reports the corresponding results and also confirms how difficult the Vox-CA test set is.
    
First, by observing the performance on our-E and our-H (our implemented following the VoxCeleb rules), the results are similar to Vox-E and Vox-H results, that demonstrate the correctness of our construction. Then, by controlling a single variable, the negative effect of cross-age is larger than the same nationality and gender matching. When we combine these elements, the performance drops dramatically with the age gap increasing. The Vox-CA provides new benchmarks for cross-age matching scenarios and hard tasks. In addition, the result of the intra-segment case is considerably lower than other test sets. It indicates that the verification of intra-segment pairs is too easy to affect the real system performance. 

\subsection{Experimental results of AISR}
Table.\ref{tab:daal_result} presents the performance of our proposed methods and related methods on different test sets. First, we compare different metric learning methods, Softmax and ArcFace. We can find that the ArcFace outperforms its counterpart, especially in cross-age scenarios. Besides, by comparing the results on cross-age test sets, we can observe that the verification performance degrades significantly with age-gap increasing.
Using the ArcFace based system, we evaluate some adversarial learning and decorrelated methods. The model with GRL or Age Residual has little improvement on cross-age scenarios. The limitation of both methods is that operations are performed on the embedding level. Since the embedding is a compact representation vector generated by the encoder layer, resulting in limited operational margins, thus the improvement is slight. In contrast with these methods, the $z_{age}$ of ADAL are extracted from high-level feature maps, and age information also can be further eliminated by the age adversarial learning classifier. 
In contrast to the baseline system, the ADAL achieves 10\% relative improvement on the Vox-CA20 test set. The larger the age gap, the better the results. In addition, the result indicates that the baseline model only mounted with ARE module is still better than other methods. 

\vspace{-0.1cm}
\section{Conclusions}
\label{sec.conclu}
This paper adopts the face age estimation method to mine the cross-age speaker verification scenario data and propose the Vox-CA as a new benchmark of CASV. The results indicate the negative effect of the cross-age matching and how difficult the Vox-CA is. In addition,  an age decoupling adversarial learning module is proposed to learn age-invariant speaker representation. Finally, the proposed method achieves 10\% relative error reduction over the baseline system on Vox-CA20. 

\bibliographystyle{IEEEtran}

\bibliography{mybib}

\begin{thebibliography}{10}
\providecommand{\url}[1]{#1}
\csname url@samestyle\endcsname
\providecommand{\newblock}{\relax}
\providecommand{\bibinfo}[2]{#2}
\providecommand{\BIBentrySTDinterwordspacing}{\spaceskip=0pt\relax}
\providecommand{\BIBentryALTinterwordstretchfactor}{4}
\providecommand{\BIBentryALTinterwordspacing}{\spaceskip=\fontdimen2\font plus
\BIBentryALTinterwordstretchfactor\fontdimen3\font minus
  \fontdimen4\font\relax}
\providecommand{\BIBforeignlanguage}[2]{{%
\expandafter\ifx\csname l@#1\endcsname\relax
\typeout{** WARNING: IEEEtran.bst: No hyphenation pattern has been}%
\typeout{** loaded for the language `#1'. Using the pattern for}%
\typeout{** the default language instead.}%
\else
\language=\csname l@#1\endcsname
\fi
#2}}
\providecommand{\BIBdecl}{\relax}
\BIBdecl

\bibitem{xvector}
D.~Snyder, D.~Garcia-Romero, G.~Sell, D.~Povey, and S.~Khudanpur,
  ``x-{vectors}: {Robust} {DNN} {Embeddings} for {Speaker} {Recognition},'' in
  \emph{Proc. {ICASSP}}, 2018, pp. 5329--5333.

\bibitem{ecapatdnn}
D.~Desplanques, J.~Thienpondt, and K.~Demuynck, ``{ECAPA-TDNN: Emphasized
  Channel Attention, Propagation and Aggregation in TDNN Based Speaker
  Verification},'' in \emph{Proc. Interspeech}, 2020, pp. 3830--3834.

\bibitem{ftdnn}
D.~Povey, G.~Cheng, Y.~Wang, K.~Li, H.~Xu, M.~Yarmohammadi, and S.~Khudanpur,
  ``{Semi-Orthogonal Low-Rank Matrix Factorization for Deep Neural Networks},''
  in \emph{Proc. Interspeech}, 2018, pp. 3743--3747.

\bibitem{cai_exploring_2018}
W.~Cai, J.~Chen, and M.~Li, ``Exploring the {Encoding} {Layer} and {Loss}
  {Function} in {End}-to-{End} {Speaker} and {Language} {Recognition}
  {System},'' in \emph{Proc. Speaker Odyssey}, 2018, pp. 74--81.

\bibitem{sphereface}
W.~Liu, Y.~Wen, Z.~Yu, M.~Li, B.~Raj, and L.~Song, ``{Sphereface: Deep
  hypersphere embedding for face recognition},'' in \emph{Proc CVPR}, 2017, pp.
  212--220.

\bibitem{arcface}
J.~{Deng}, J.~{Guo}, N.~{Xue}, and S.~{Zafeiriou}, ``{ArcFace: Additive Angular
  Margin Loss for Deep Face Recognition},'' in \emph{Proc. CVPR}, 2019, pp.
  4685--4694.

\bibitem{modadp_sv}
W.~Mistretta and K.~Farrell, ``{Model adaptation methods for speaker
  verification},'' in \emph{Proc. ICASSP}, 1998, pp. 113--116.

\bibitem{sv_longtime_aging}
F.~Kelly, A.~Drygajlo, and N.~Harte, ``{Speaker verification with long-term
  ageing data},'' in \emph{Proc. IAPR International Conference on Biometrics},
  2012, pp. 478--483.

\bibitem{effect_ca_sv}
F.~Kelly and N.~Harte, ``{Effects of Long-Term Ageing on Speaker
  Verification},'' in \emph{Biometrics and ID Management}, 2011, pp. 113--124.

\bibitem{voiceaging_sr}
Y.~Matveev, ``The problem of voice template aging in speaker recognition
  systems,'' in \emph{International Conference on Speech and Computer}, 2013,
  pp. 345--353.

\bibitem{roleage_sv}
Y.~Lei and J.~H. Hansen, ``The role of age in factor analysis for speaker
  identification,'' in \emph{Tenth Annual Conference of the International
  Speech Communication Association}, 2009.

\bibitem{AISE_diarizaion}
S.~S. Xu, M.-W. Mak, K.~H. Wong, H.~Meng, and T.~C.~Y. Kwok, ``{Age-Invariant
  Speaker Embedding for Diarization of Cognitive Assessments},'' in \emph{Proc.
  ISCSLP}, 2021, pp. 1--5.

\bibitem{age_nist_ivector}
S.~O. Sadjadi, S.~Ganapathy, and J.~W. Pelecanos, ``{Speaker age estimation on
  conversational telephone speech using senone posterior based i-vectors},'' in
  \emph{Proc. ICASSP}, 2016, pp. 5040--5044.

\bibitem{age_nist_e2e}
P.~Ghahremani, P.~S. Nidadavolu, N.~Chen, J.~Villalba, D.~Povey, S.~Khudanpur,
  and N.~Dehak, ``{End-to-end Deep Neural Network Age Estimation.}'' in
  \emph{Proc. Interspeech}, 2018, pp. 277--281.

\bibitem{age_timit}
T.~Gupta, D.-T. Truong, T.~T. Anh, and C.~E. Siong, ``Estimation of speaker age
  and height from speech signal using bi-encoder transformer mixture model,''
  \emph{arXiv preprint arXiv:2203.11774}, 2022.

\bibitem{age_ml}
M.~Li, K.~J. Han, and S.~Narayanan, ``Automatic speaker age and gender
  recognition using acoustic and prosodic level information fusion,''
  \emph{Computer Speech \& Language}, pp. 151--167, 2013.

\bibitem{age_ivector_ml}
P.~G. Shivakumar, M.~Li, V.~Dhandhania, and S.~S. Narayanan, ``Simplified and
  supervised i-vector modeling for speaker age regression,'' in \emph{Proc.
  ICASSP}, 2014, pp. 4833--4837.

\bibitem{timit}
J.~S. Garofolo, L.~F. Lamel, W.~M. Fisher, J.~G. Fiscus, and D.~S. Pallett,
  ``Darpa timit acoustic-phonetic continous speech corpus cd-rom. nist speech
  disc 1-1.1,'' \emph{NASA STI/Recon technical report n}, p. 27403, 1993.

\bibitem{fairvoice}
G.~Fenu, M.~Marras, G.~Medda, and G.~Meloni, ``{Fair Voice Biometrics: Impact
  of Demographic Imbalance on Group Fairness in Speaker Recognition},'' in
  \emph{Proc. Interspeech 2021}, 2021, pp. 1892--1896.

\bibitem{vox1}
A.~Nagrani, J.~Chung, and A.~Zisserman, ``Voxceleb: {A} {Large}-{Scale}
  {Speaker} {Identification} {Dataset},'' in \emph{Proc. Interspeech}, 2017,
  pp. 2616--2620.

\bibitem{vox2}
J.~Chung, A.~Nagrani, and A.~Zisserman, ``Voxceleb2: {Deep} {Speaker}
  {Recognition},'' in \emph{Proc. Interspeech}, 2018.

\bibitem{agevoxceleb}
N.~Tawara, A.~Ogawa, Y.~Kitagishi, and H.~Kamiyama, ``{Age-VOX-Celeb:
  Multi-Modal Corpus for Facial and Speech Estimation},'' in \emph{Proc.
  ICASSP}, 2021, pp. 6963--6967.

\bibitem{vox_enrichment}
K.~Hechmi, T.~N. Trong, V.~Hautamäki, and T.~Kinnunen, ``{Voxceleb Enrichment
  for Age and Gender Recognition},'' in \emph{ASRU}, 2021, pp. 687--693.

\bibitem{calfw}
T.~Zheng, W.~Deng, and J.~Hu, ``{Cross-Age LFW: A Database for Studying
  Cross-age Face Recognition in Unconstrained Environments},'' \emph{arXiv
  preprint arXiv:1708.08197}, 2017.

\bibitem{dal_aifr}
H.~Wang, D.~Gong, Z.~Li, and W.~Liu, ``{Decorrelated Adversarial learning for
  age-invariant face recognition},'' in \emph{Proc. CVPR}, 2019, pp.
  3527--3536.

\bibitem{orthogonal_decom_aifr}
Y.~Wang, D.~Gong, Z.~Zhou, X.~Ji, H.~Wang, Z.~Li, W.~Liu, and T.~Zhang,
  ``{Orthogonal Deep Features Decomposition for Age-invariant Face
  Recognition},'' in \emph{Proc. ECCV}, 2018, pp. 738--753.

\bibitem{aifr_fas}
Z.~Huang, J.~Zhang, and H.~Shan, ``{When age-invariant face recognition meets
  face age synthesis: A multi-task learning framework},'' in \emph{Proc. CVPR},
  2021, pp. 7282--7291.

\bibitem{dex}
R.~Rothe, R.~Timofte, and L.~Van~Gool, ``{Dex: Deep expectation of apparent age
  from a single image},'' in \emph{ICCV}, 2015, pp. 10--15.

\bibitem{asp_pooling}
K.~Okabe, T.~Koshinaka, and K.~Shinoda, ``{Attentive Statistics Pooling for
  Deep Speaker Embedding},'' \emph{Proc. Interspeech}, 2018.

\bibitem{grl}
Y.~Ganin and V.~Lempitsky, ``{Unsupervised Domain Adaptation by
  Backpropagation},'' in \emph{Proc. ICML}, 2015, pp. 1180--1189.

\bibitem{resnet}
K.~He, X.~Zhang, S.~Ren, and J.~Sun, ``{Deep Residual Learning for Image
  Recognition},'' in \emph{Proc. CVPR}, 2016, pp. 770--778.

\bibitem{on-the-fly}
W.~{Cai}, J.~{Chen}, J.~{Zhang}, and M.~{Li}, ``{On-the-Fly Data Loader and
  Utterance-Level Aggregation for Speaker and Language Recognition},''
  \emph{IEEE/ACM Transactions on Audio, Speech, and Language Processing}, pp.
  1038--1051, 2020.

\bibitem{musan}
D.~Snyder, G.~Chen, and D.~Povey, ``{MUSAN}: {A} {Music}, {Speech}, and {Noise}
  {Corpus},'' \emph{arXiv:1510.08484}.

\bibitem{RIR}
T.~Ko, V.~Peddinti, D.~Povey, M.~Seltzer, and S.~Khudanpur, ``A study on data
  augmentation of reverberant speech for robust speech recognition,'' in
  \emph{Proc. ICASSP}, 2017, pp. 5220--5224.

\end{thebibliography}

\end{document}